\documentclass[fleqn,usenatbib,a4paper]{mnras}

\usepackage{newtxtext,newtxmath}

\usepackage[T1]{fontenc}
\usepackage{ae,aecompl}
\usepackage{times}

\newcommand{\rd}[1]{{#1}}

\usepackage{graphicx}	
\usepackage{amsmath}	
\usepackage{amssymb}	

\usepackage{tikz}
\usepackage{pgfplots}
\usetikzlibrary{pgfplots.groupplots}

\newcommand{\psiH}{\psi_{\rm H}}
\newcommand{\rhoH}{\rho_{\rm H}}
\newcommand{\psiSH}{\psi_{\rm sNFW}}
\newcommand{\rhoSH}{\rho_{\rm sNFW}}
\newcommand{\rhoNFW}{\rho_{\rm NFW}}
\newcommand{\psiNFW}{\psi_{\rm NFW}}

\newcommand{\rv}{r_{\rm v}}
\newcommand{\Mv}{M_{\rm v}}
\newcommand{\Reff}{R_{\rm e}}
\newcommand{\rs}{r_{\rm s}}

\newcommand{\ellipE}{{\rm E}\!}
\newcommand{\ellipK}{{\rm K}\!}
\newcommand{\ellipF}{{\rm F}\!}
\newcommand{\ellipPi}{{\rm \Pi}\!}
\newcommand{\cNFW}{{c_{\rm NFW}}}
\newcommand{\cSH}{{c_{\rm sNFW}}}

\title[The Super-NFW model]{The Super-NFW model: an analytic dynamical model for cold dark matter haloes and elliptical galaxies}

\author[E. Lilley et al.]{Edward J. Lilley,$^{1}$\thanks{E-mail:
    ejl44,nwe,jls@cam.ac.uk} N. Wyn Evans$^{1}$, Jason
  L. Sanders$^{1}$ \\
$^{1}$Institute of Astronomy, Madingley Rd, Cambridge, CB3 0HA, UK\\
}
\date{Accepted XXX. Received YYY; in original form ZZZ}

\pubyear{2015}

\begin{document}
\label{firstpage}
\pagerange{\pageref{firstpage}--\pageref{lastpage}}
\maketitle

\begin{abstract}
An analytic galaxy model with $\rho \sim r^{-1}$ at small radii and
$\rho \sim r^{-3.5}$ at large radii is presented. \rd{The asymptotic density fall-off is slower than the Hernquist model, but faster than the Navarro-Frenk-White (NFW) profile for dark matter haloes, and so in accord with recent evidence from cosmological simulations. The model provides the zeroth-order term in a biorthornomal basis function expansion, meaning that axisymmetric, triaxial and lopsided distortions can easily be added (much like the Hernquist model itself which is the zeroth-order term of the Hernquist-Ostriker expansion)}.  The properties of the \rd{spherical} model, including analytic distribution functions which are either isotropic, radially anisotropic or tangentially anisotropic, are discussed in some detail. The analogue of the mass-concentration relation for cosmological haloes is provided.
\end{abstract}


\section{Introduction}

Analytic galaxy profiles provide simplicity for the modelling of
galaxy components as well as insight into the dynamics of more
realistic galaxies. For instance, the \citet{He90} galaxy model has
the potential-density pair
 \begin{eqnarray}
 \psiH &=& {GM\over r+b}, \nonumber \\
 \rhoH &=& {Mb\over 2\pi}{1\over r (r+b)^3},
 \label{eq:hernquist}
 \end{eqnarray}
where $b$ is a \rd{parameter with the dimensions of length}. It is related to the effective radius (or the radius of contour of surface density enclosing  half the mass) by $\Reff = 1.815 b$. This profile is often used to model dark haloes which are believed to have the universal or \citet[][henceforth NFW]{Na97} form
\begin{eqnarray}
\psiNFW &=&  {4 \pi G \rho_0 \rs^3\over r}\log(r+\rs),\nonumber\\
\rhoNFW &=& {\rho_0 \rs^3 \over r (r + \rs)^2},
\label{eq:nfw}
\end{eqnarray}
where $\rs$ is a scalelength related to the virial radius. The asymptotic fall-off of the density in the Hernquist model is $\rhoH \sim r^{-4}$, so it avoids the defect of infinite mass which afflicts the NFW halo with $\rhoNFW \sim r^{-3}$. Additionally, the Hernquist model is often used to represent bulges and elliptical galaxies as it follows the \citet{deV53} profile to a good approximation.

In this paper, we introduce the {\it \rd{super-NFW (or sNFW) model}},
which has density and potential pair:
\begin{eqnarray}
\psiSH &=& {GM \over r + a  + \sqrt{a}\sqrt{r + a}}  \nonumber\\
\rhoSH &=& {3M\sqrt{a} \over 16 \pi}{1 \over r(r+a)^{5/2}},
\label{eq:sh}
\end{eqnarray}
where $a$ is related to the effective radius by $\Reff = 5.478a$. 

Why is it `\rd{super-NFW}'? \rd{The model provide a good match to cosmological
  haloes, but it has finite mass as the density falls off more slowly
  like $\rhoSH \sim r^{-3.5}$. This is slower than the Hernquist model,
  but faster than the NFW. Recent work on the splashback radius~\citep{Di17} 
  suggests that the density of cosmological haloes drops more rapidly than 
  NFW, but slower than Hernquist, at large radii. There are of course
  other models in the literature that have $\rho \propto r^{-1}$ at
  the centre and have an asymptotic fall off with logarithmic gradient
  between $-3$ and $-4$ \citep[e.g.][]{An06,An13}. The super-NFW
  model however has another special property -- it is the zeroth-order
  term of a biorthogonal expansion~\citep[][henceforth Paper II]{Li17}. The 
  potential-density pairs of arbitrarily distorted sNFW models are, therefore, 
  straightforward to construct. This is important as cosmological dark haloes 
  show many deviations from spherical symmetry, and therefore a spherical model 
  is only good to crudest order.}

\rd{The sNFW model is part of the general double-power law family
  investigated by \citet{Zh96} and subsequently studied in detail by
  \mbox{\cite{An13}}, namely
\begin{equation}
\rho(r) =  {C \over r^\gamma(1 + r^{1/\alpha})^{(\beta-\gamma)\alpha}},
\end{equation}
where $C$ is a normalisation constant. In Zhao's notation, the
\rd{sNFW} model has $(\alpha,\beta,\gamma) = (1,3.5,1)$. Zhao was the first to note that the potential is simple. The \rd{sNFW} model is also the $b=7/2$ member of the generalized NFW family
\begin{equation}
\rho(r) =  {C \over r (1 + r)^{b-1}},
\end{equation}
studied in \citet{Ev06}, who give asymptotic results for the dynamical quantities for the whole family. In general, however, biorthogonal basis sets for these entire families of models are not known, meaning that it is hard to build axisymmetric or triaxial analogues. This, however, is not the case for the sNFW model, as we show in our accompanying Paper II.}

\begin{figure}
  \centering
  \includegraphics[width=0.8\columnwidth]{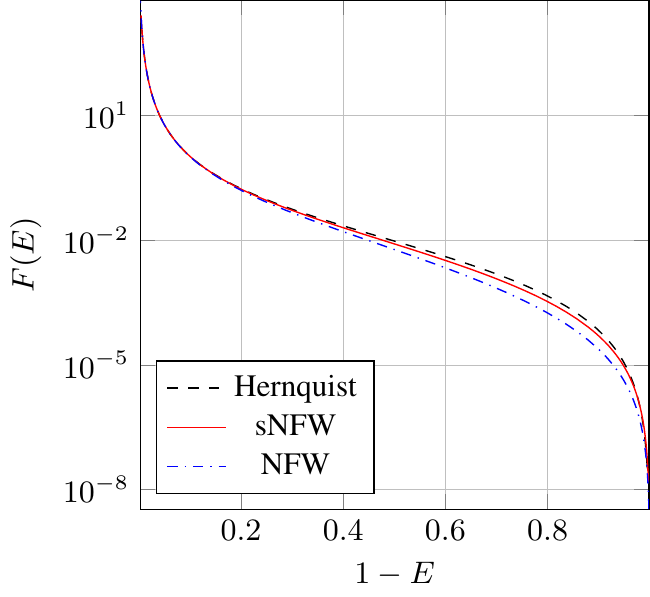}
  \caption{\rd{The isotropic DFs $F(E)$ of the \rd{sNFW} model (red) compared to the DFs of the Hernquist model (black) and NFW model (blue). The three models have the same central value of the potential and the same halo scalelength $\rs$.}}
\label{fig:DFs}
\end{figure}
\begin{figure}
	\includegraphics[width=0.8\columnwidth]{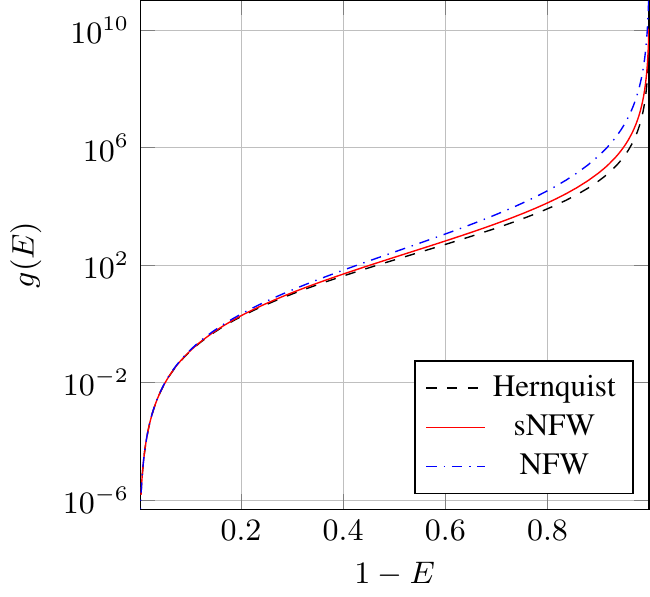}
    \caption{\rd{The density of states $g(E)$ of the \rd{super-NFW} model (red) compared to the Hernquist (black) and NFW (blue) models.}}
    \label{fig:dos}
\end{figure}

\section{The Model}
\subsection{Isotropic distribution functions}

Let us use units in which $G=a=1$ and $M=2$, so that the central value of the \rd{sNFW} potential is $\psiSH(0) =1$. The enclosed mass is
\begin{equation}
M(r) =  2 - {2+3r \over (1+r)^{3/2}},
\end{equation}
so that the half-mass radius is $r_{\rm 1/2}  = 7.29086$.

The potential $\psi(r)$ can be inverted simply by setting $x = \sqrt{1
  + r}$ and solving the resulting quadratic in $x$; the form
of $r(\psi)$ is then
\begin{equation}
r(\psi) =  {4- \psi -\sqrt{\psi(8+\psi)}\over 2 \psi},
\end{equation}
so that the density $\rho$ can be expressed as
\begin{equation}
\rho(\psi) =
{3\psi\,\left(4-\psi+\sqrt{\psi(8+\psi)}\right)\,
\left(\psi + \sqrt{\psi(8+\psi)}\right)^5 
\over 2^{16} \pi (1-\psi)}.
\end{equation}

The isotropic distribution function (DF) is then given by \citet{Edd16} as
\begin{equation}
F(E) = {1\over \sqrt{8}\pi^2}{d\over dE}\left[ \int_0^E{d\rho\over d\psi} {d\psi\over \sqrt{E-\psi}}\right],
\end{equation}
This can be evaluated exactly as
\begin{eqnarray}
F(E) &=& {3\over 7\cdot 2^{10}\pi^3(8+E)(1-E)^2}\Bigl[
252\tfrac{8+E}{\sqrt{2(1-E)}} \arcsin \sqrt{E}\nonumber \\
&+& P_1(E)\sqrt{\tfrac{E}{2}} + P_2(E)\:\ellipE\left(-\tfrac{E}{8}\right)
+ P_3(E)\:\ellipK\left(-\tfrac{E}{8}\right) \\ 
&+& 189(8+E)\:\ellipPi\left(E,-\tfrac{E}{8}\right)\Bigr].\nonumber
\end{eqnarray}
where E, K and $\Pi$ are complete elliptic integrals and the $P_i$ are
the polynomials given in Appendix A. This is more complicated than the
isotropic DF of the Hernquist Model, but simpler than the isotropic DF
of the NFW model, which was first numerically constructed \citep{Wi00,
  Lo01} and then analytically derived by~\cite{Ev06}. 
  
In Fig.~\ref{fig:DFs}, we show the isotropic DF of this model against
that of the Hernquist \rd{and NFW models}. To compare all three models, we 
use the \textit{halo scalelength} $\rs$, which is defined as the radius at 
which the logarithmic slope of the density attains the isothermal value, that is
\begin{equation}\label{eq:scalelength}
\left.\frac{d \log{\rho}}{d \log{r}}\right|_{r=\rs} = -2.
\end{equation}
For the Hernquist model we find $\rs = b/2$; for \rd{sNFW} $\rs =
2a/3$, and for NFW the initial choice of $\rs$ already satisfies this
property.
As the form of the cusp is the same at small radii ($\rho \sim 1/r$),
so the DFs of all three models diverge like
$(1-E)^{-5/2}$ as $E \rightarrow 1$. However, for stars close to the
binding energy ($E \rightarrow 0$), the Hernquist DF behaves like
$E^{5/2}$, the \rd{sNFW} like $E^2$ and the NFW like $E^{3/2}$.
The density of states is 
\begin{equation}
g(E)  = (4 \pi)^2 \int_0^{r_E} r^2 \sqrt{ 2 (E-\psi)} dr
\end{equation}
where $r_E$ is the maximum radius of orbit with energy $E$
~\citep{BT}. After some calculation, we obtain
\begin{eqnarray}
  g(E) = \frac{4\sqrt{2}\pi^2}{3 E^2} \left\{ 12\arccos{\left(\sqrt{E}\right)}\frac{1+E}{\sqrt{E}} + 12(1-E)^{3/2}(2E+3) \right. \\
    + \left. \sqrt{8+E}\left[(2E^2+7E-16)\:\ellipE\left(\phi,\kappa^2\right) + E(1-2E)\:\ellipF\left(\phi,\kappa^2\right) \right] \right\},
\end{eqnarray}
where F and E are incomplete elliptic integrals\footnote{Note that we
  we use the \textit{Mathematica} convention for the arguments of the
  elliptic functions, so that
  $\mathrm{E}(\phi,m)=\int^\phi_0\mathrm{d}\theta\,\sqrt{1-m\sin^2\theta}$.},
with argument $\phi \equiv \arctan\sqrt{(1-E)/E}$ and modulus
$\kappa^2 \equiv 8/(8+E)$.
The density of states for the \rd{sNFW} model is compared to that of
the Hernquist \rd{and NFW models} in Fig.~\ref{fig:dos}.

\begin{figure}
	\includegraphics[width=0.8\columnwidth]{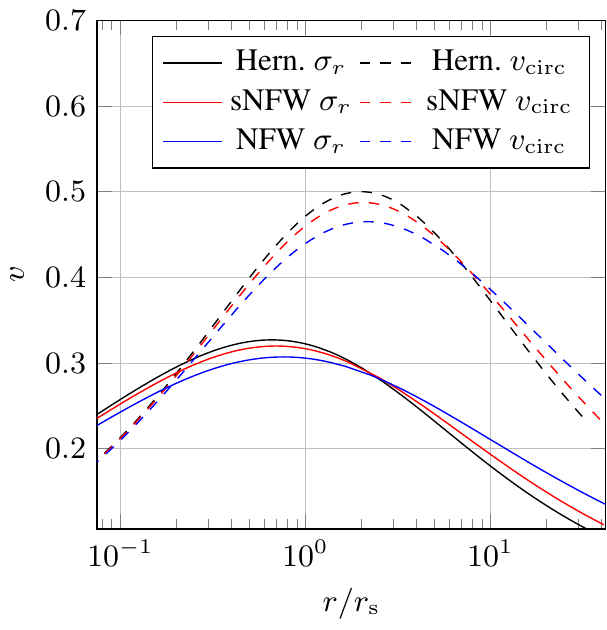}
    \caption{The velocity dispersion (full) and rotation curve (dotted) of the \rd{sNFW} model (red) compared to the Hernquist (black) \rd{and NFW (blue) models plotted against radius (in units
of $\rs$)}. Note the peaks of the rotation curve and velocity dispersion of the models are comparable, but the decline for the Hernquist model takes place more quickly than for the
\rd{sNFW} and NFW models.}
\label{fig:vdisps}
\end{figure}

The isotropic velocity dispersion is 
\begin{eqnarray}
\langle v_r^2 \rangle &=& {1\over 6r(1+r)} \Bigl[6 - 9 r - 176 r^2 -406 r^3 - 350 r^4 - 105 r^5 \nonumber \\
&-&\sqrt{1+r}\left(6-12r -88r^2-120r^3 -48r^4\right) \\
&+& 3r^2(1+r)^{7/2}\left(35{\rm arccosh}\sqrt{r} - 16\log\left( 1+ \tfrac{1}{r}\right) \right).\nonumber
\end{eqnarray}
The circular velocity curve (or rotation curve) is 
\begin{equation}
v_{\rm circ}^2 =  {2(1+r)^{3/2} -3r -2 \over r(1+r)^{3/2}}.
\end{equation}
In Fig.~\ref{fig:vdisps}, the velocity dispersion profile and the
circular velocity curve are shown for this model, as well as for the
Hernquist \rd{and NFW models}. Both the velocity
dispersion and the rotation curve of the \rd{sNFW} model have the desirable feature that they fall off much more slowly than for the Hernquist model. This is useful in modelling elliptical galaxies. For example, \citet{Ge01} found that the circular
velocity profiles of giant ellipticals are flat to within 10 to 20 per cent between $0.2 \Reff$ to at least $2\Reff$, independent of luminosity.

\begin{figure}
	\includegraphics[width=\columnwidth]{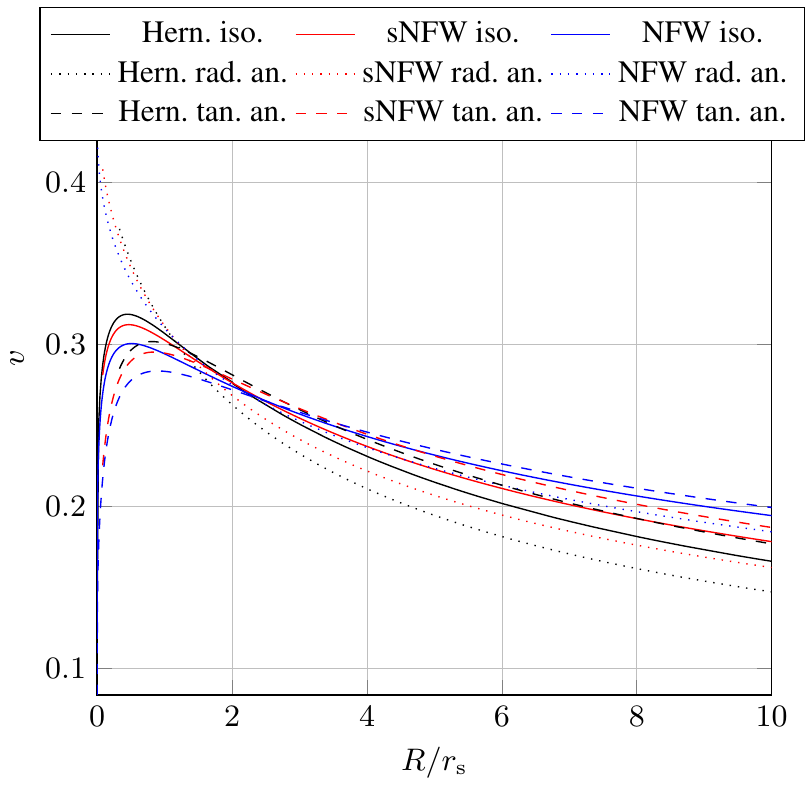}
    \caption{The line of sight velocity dispersions for the \rd{sNFW}, Hernquist \rd{and NFW} models plotted against projected distance
(in units of $r_s$). Full lines are the isotropic model, dotted radially anisotropic ($\beta = \tfrac{1}{2}$) and dashed tangentially anisotropic ($\beta = -\tfrac{1}{2}$).}
\label{fig:obs}
\end{figure}

\subsection{Anisotropic distribution functions}

Analyses of kinematic data suggest that most elliptical galaxies are
close to isotropic~\citep{Ge01}. Anisotropy is usually parametrized
via
\begin{equation}
\beta = 1 - {\langle v_\theta^2 \rangle + \langle v_\phi^2 \rangle \over 2\langle v_r^2 \rangle},
\end{equation}
where angled brackets denote averages over the DF. \citet{Ge01} find
that $-0.5 \lesssim \beta \lesssim 0.5$ in their study of giant
ellipticals. Mild radial anisotropy is most common, though some
tangential anisotropic elliptical galaxies are known. We now develop
two models that bracket the range of relevant anisotropies.

The DF of a spherical system with constant anisotropy is
\begin{equation}
F(E,L)=L^{-2\beta}f_E(E).
\label{eq:ansatz}
\end{equation}
The unknown function $f_E(E)$ can be recovered from the integral
inversion formula~\citep[see e.g.,][]{Wi99,Ev06}
\begin{equation}
f_E(E)=
\frac{2^\beta(2\pi)^{-3/2}}{\Gamma(1-\lambda)\Gamma(1-\beta)}\
\frac{d}{dE}\!\int_0^E\frac{d\psi}{(E-\psi)^\lambda}
\frac{d^nh}{d\psi^n},
\label{eq:disint}
\end{equation}
where $h=r^{2\beta}\rho$ is expressed as a function of $\psi$, and
$n=\lfloor(3/2-\beta)\rfloor$ and $\lambda=3/2-\beta-n$ are the
integer floor and the fractional part of $3/2-\beta$. This includes
the \citet{Edd16} formula for the isotropic DF as a special case
($\beta=0$).

For the radially anisotropic model when $\beta = 1/2$, the expression
for the DF reduces to
\begin{equation}
F(E,L)=\frac1{2\pi^2}\frac{1}{L}
\left.\frac{dh}{d\psi}\right|_{\psi=E} = {f_E(E)\over L},
\label{eq:dishalf}
\end{equation}
which for our model becomes
\begin{equation} f_E(E) =
\frac{15 \left( E + \sqrt{E(E+8)} +4\right) \left(E +\sqrt{E (E\!+\!8)}\right)^4}{2^{13} \pi \sqrt{E(E\!+\!8)}}.
\end{equation}
The radial velocity dispersion is
\begin{eqnarray}
\langle v_r^2 \rangle &=& {1\over 6r(r+1)} \biggl[ 12\left(\sqrt{r+1}-1\right) + 10 r^3 \left(-6 r+3 \sqrt{r+1}-20\right)\nonumber\\
&+&75r^2\sqrt{r+1}-232 r +55\sqrt{r+1}-104r \\
&+&30 r (r+1)^{7/2} \left(\log\left(\frac{r}{r+1}\right) + 2\:\text{arccsch}\sqrt{r}\right) \Biggr].\nonumber
\end{eqnarray}
The analogous radially anisotropic ($\beta = 1/2$) DFs for the
Hernquist model is very simple and was discovered by \citet{Ba02} and
\citet{Ev05}.

For the tangentially anisotropic model when $\beta = -1/2$, the
expression for the DF further reduces to
\begin{equation}
F(E,L)=\frac{L}{2\pi^2}
\left.\frac{d^2h}{d\psi^2}\right|_{\psi=E} = L f_E(E) ,
\label{eq:dismhalf}
\end{equation}
where 
\begin{equation}
f_E(E) ={3E^{3/2}(E\!+\!\sqrt{E(E+8)})^4 (36\!-\!19E\!-\!5E^2\!+\!\sqrt{E(E\!+\!8)}(9\!-\!3E))\over 2^{12}\pi^3 \sqrt{E\!+\!8}(1\!-\!E)(E\!+\!\sqrt{E(8\!+\!E)}\!-\!4)^2}
\end{equation}
whilst the second moment is
\begin{eqnarray}
\langle v_r^2 \rangle &=&{1 \over 12 r(r+1)} \Biggl[
-15 r^3 (r+1)^{7/2} (21 \text{arccsch} \sqrt{r} + 8 \log {r\over r+1})\nonumber \\
&+& 8(\sqrt{r\!+\!1}\!-\!1)\!+\!5r^2 (6 \sqrt{r\!+\!1}\!-\!7)+ 44r^3 (5 \sqrt{r\!+\!1}-12)\nonumber\\
&-&3 r^4(406 -100 \sqrt{r+1})-
5r^5(21 r-8 \sqrt{r\!+\!1}+70)\Biggr].
\end{eqnarray}

Fig.~\ref{fig:obs} shows the line of sight velocity dispersions for
the \rd{sNFW} model for the three choices of anisotropy ($\beta =
\tfrac{1}{2},0$ and $-\tfrac{1}{2}$). The equivalent results for the
Hernquist \rd{and NFW models} are also shown. As expected, radial
anisotropy leads to an enhancement of the line of sight dispersion
near the centre (where the dotted curves lie above the full
curves). Tangential anisotropy causes the line of sight dispersions to
be enhanced above the isotropic case in the outer parts (where the
dashed curves lie above full curve). Note that the line of sight
dispersion profiles of the \rd{sNFW} model show a more gradual decline
with distance than the Hernquist model. This is in good accord with
the data on nearly round elliptical galaxies, which show slow declines
out to $\approx 2 \Reff$ ~\citep{Kr00}.

Overall, the \rd{sNFW} model gives somewhat more complicated
expressions for quantities (such as the DFs) than the Hernquist
model. The pay-back is that the density profile falls off more slowly
($\rhoSH \sim r^{-3.5}$) and so the rotation curve and velocity
dispersion profiles are much flatter. This is much more like the
observed behaviour of elliptical galaxies and dark matter haloes.

\section{Comparisons}

\begin{figure}
	\includegraphics[width=0.8\columnwidth]{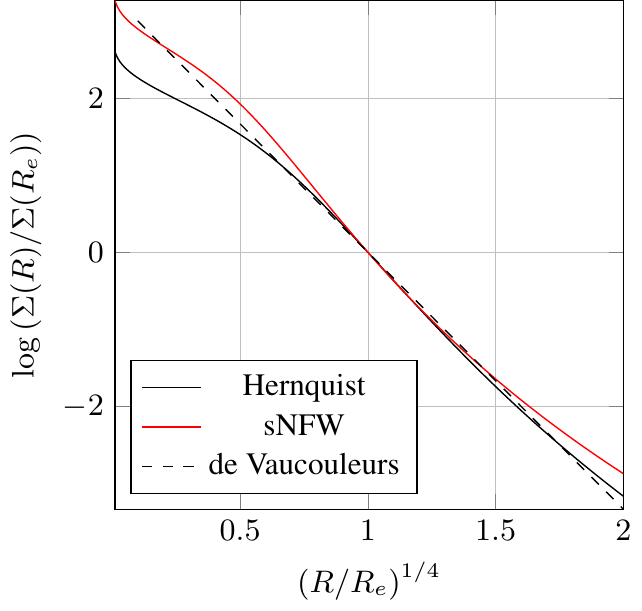}
    \caption{The surface density of the \rd{sNFW} model (red), the Hernquist model (black) and the de Vaucouleurs profile (dashed).} 
    \label{fig:proj}
\end{figure}

\subsection{Sersic and de Vaucouleurs Profile}

Traditionally, the light profiles of elliptical galaxies and bulges
have been fit by the \citet{deV53} profile, which is
\begin{equation}
\log_{10} \left( {\Sigma(R)\over \Sigma(\Reff)} \right) =  - 3.331  \left[ \left( {R\over \Reff}\right)^{1/4} -1\right].
\end{equation}
Here, $\Reff$ is the effective radius, or the radius of the isophote
that encloses half the luminosity. \citet{Ca93} examined the
photometric profiles of a large sample of elliptical galaxies and
argued for use of the more general \citet{Se68} law
\begin{equation}
\log_{10} \left( {\Sigma(R) \over \Sigma(\Reff)} \right) \approx 0.8686(0.1635-n) \left[\left({R\over\Reff}\right)^{1/n}-1\right],
\end{equation}
where the Sersic index $n$ satisfies $2 \lesssim n \lesssim 10$. The
de Vaucouleurs profile is the case $n=4$, whilst $n=1$ corresponds to
an exponential. The photometric profiles of bulges were studied by
\citet{An95}, who found that bulges of S0s are well-fit by a de
Vaucouleurs profile, whilst bulges of late-type spirals are better fit
by an exponential.

To compare the \rd{sNFW} model against these photometric laws, we must
first compute its projected density. This is
\begin{equation}
\Sigma(R) = \frac{(R+1) (R+3) \: \ellipK\left(\frac{R-1}{2 R}\right)-8 R \: \ellipE\left(\frac{R-1}{2 R}\right)}{2^{3/2} \pi  \sqrt{R} \left(R^2-1\right)^2}.
\end{equation}
The half-light or effective radius is $\Reff =1.81527$. Assuming a
constant mass-to-light ratio, Fig.~\ref{fig:proj} shows the surface
brightness of the de Vaucouleurs profile, together with the Hernquist
and \rd{sNFW} models between $0.1$ and $2 \Reff$. The \rd{sNFW} model
is a better global fit than the Hernquist model to the de Vaucouleurs
profile. Beyond $\approx 5 \Reff$, however, the \rd{sNFW} surface
density falls off rather more slowly than both the Hernquist and de
Vaucouleurs profiles.

More formally, we can fit the projected densities of both the
Hernquist and \rd{sNFW} models between $0.1$ and $2\Reff$ to the
Sersic profiles. The Hernquist model gives $n = 3.388$ and the
\rd{sNFW} $n=4.200$. Given the range of properties of elliptical
galaxies and bulges, both profiles are useful. The Hernquist model is
a better match to $n \approx 3$ Sersic profiles, whilst the \rd{sNFW}
is a better match to $n \approx 4$ (de Vaucouleurs profile).

\subsection{Numerical halo fitting}

\begin{figure}
\centering
\includegraphics{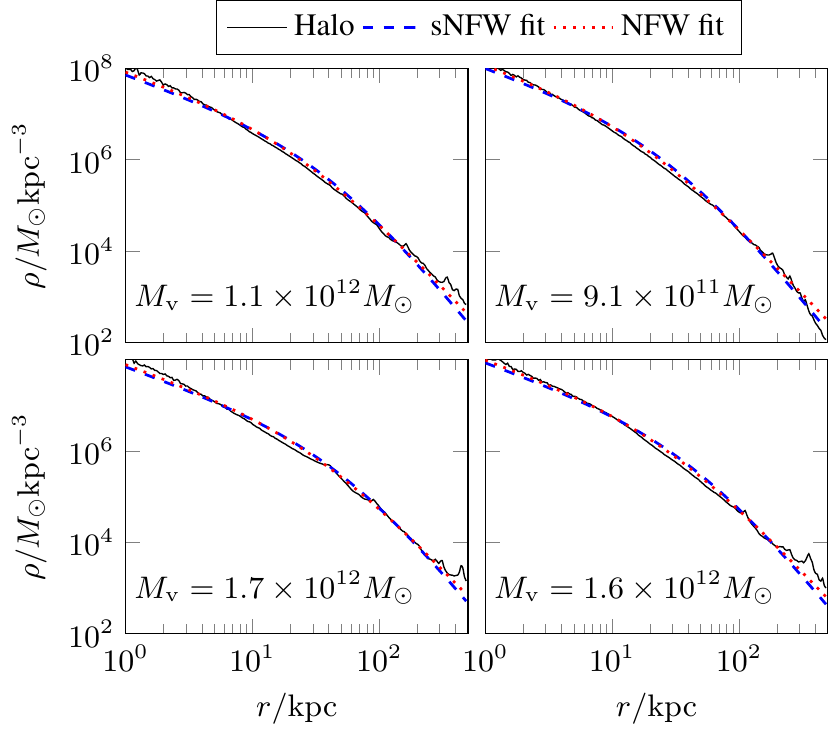}
\caption{Fits of the \rd{sNFW} and NFW models to the radial density
  profile of four dark matter haloes extracted from a cosmological
  $N$-body simulation. The numerical halo data is binned
  logarithmically, and the virial mass of each halo is
  inlaid.}\label{fig:halo_fits}
\end{figure}

\begin{figure}
\centering
\includegraphics{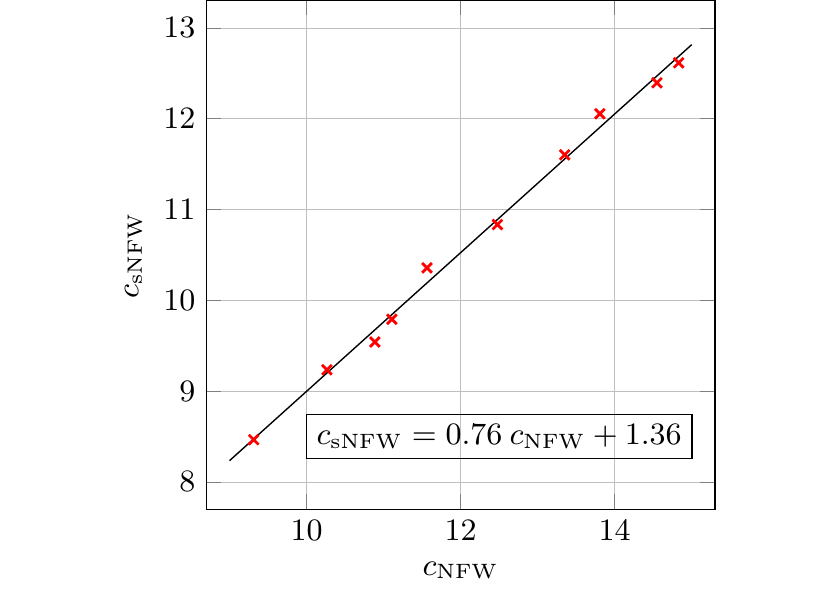}
  \caption{Concentrations for 10 numerical halos as fitted
    by the \rd{sNFW} and NFW models. The line of best fit is shown.}\label{fig:conc_conc}
\end{figure}

Dark matter halos are often parametrised in terms of their virial mass
$\Mv$, virial radius $\rv$ and concentration $c$~\citep{Di15}. The
virial mass is the mass contained within a spherical shell of radius
$\rv$, that is $\Mv = M(\rv)$. Once a particular model is chosen to
fit the halo, the length scale is parametrised using the concentration
parameter $c = \rv/\rs$, where $\rs$ is the halo scalelength as
defined in Eq.~\eqref{eq:scalelength}.
For the NFW model, in the notation of equation \eqref{eq:nfw}, we have
\begin{equation}
\rhoNFW = \frac{\Mv}{4\pi} \: \left(\log(1+\cNFW) - \frac{\cNFW}{1+\cNFW}\right)^{-1} \: \frac{1}{r \left(r + \rv/\cNFW\right)^2}.
\end{equation}
The analogous definition for the \rd{sNFW} model is
\begin{equation}
\rhoSH = \frac{3^{3/2}\Mv}{8\pi} \: \left(1 - \frac{1+\cSH}{\left(1 + 2\cSH/3\right)^{3/2}}\right)^{-1} \: {(\rv/\cSH)^{1/2}\over {r\left(2r + 3\rv/\cSH\right)^{5/2}}}.
\end{equation}
We have fitted both the NFW and \rd{sNFW} model to ten
numerically-constructed dark matter haloes extracted from cosmological
simulations~\rd{(for more details on their provenance, see Paper
  II)}. Four of these fits are shown in Fig.~\ref{fig:halo_fits}, and
the relation between the derived concentration parameters for the two
models, along with a best fit line, is shown in
Fig.~\ref{fig:conc_conc}. We notice that the \rd{sNFW} model does at
least as good a job as the NFW profile in fitting the shapes of these
ten haloes. The concentration $\cSH$ of the best-fit \rd{sNFW} model
is related to that of the best-fit NFW model via
\begin{equation}
\cSH = 1.36 + 0.76\cNFW
\end{equation}
This gives an easy way to transform the mass-concentration relations
for NFW models, given for example in equations (8) and (9) of
\citet{Du14}, to provide a cosmologically-inspired sequence of
\rd{sNFW} dark haloes.


\section{Conclusions}

We have introduced the \rd{super-NFW} (\rd{sNFW}) model. This is a
potential-density pair useful for representing \rd{spherical} bulges,
elliptical galaxies and dark haloes. The radial force component is very
simple, namely
\begin{equation}
F_r = \nabla \psi =  
\frac{\sqrt{a} GM}{r^2 \sqrt{a+r}}+\frac{\sqrt{a} GM}{2 r (a+r)^{3/2}}-\frac{GM}{r^2}
\end{equation}
This should make the model useful for $N$-body work. 

The density of the \rd{sNFW} model falls like $\rhoSH \sim r^{-1}$ near
the centre, and like $\rhoSH \sim r^{-3.5}$ on the outskirts. This gives it two important advantages over competitors like the Hernquist model: First, it is a better match to the de Vaucouleurs profile in the inner parts, so it is useful for modelling the light profiles of elliptical galaxies and bulges. Its density falls off somewhat more slowly than the Hernquist model ($\rhoH \sim r^{-4}$), which makes it a better match to the line of sight velocity dispersion profiles of ellipticals at large radii. Secondly, the asymptotic density fall-off is closer to the density profile of numerically-constructed haloes, which approximately follow the Navarro-Frenk-White (NFW) form and have $\rhoNFW \sim r^{-3}$ at large radii.  The advantage of using a \rd{sNFW} model rather than an NFW model is that we have provided a suite of DFs (isotropic, radially and tangentially anisotropic) for the \rd{sNFW} model, whereas the DFs of the NFW model are not elementary~\citep{Ev06}.

\rd{There are of course many other models with central density cusps
  like $\rho \sim r^{-1}$ and with asymptotic decays between $\rho
  \sim r^{-3}$ and $\rho \sim r^{-4}$ \citep[e.g.,][]{An06} or halo models with finite mass \citep[e.g.,][]{Na04,Zh05}. In particular, \citet{An13} provide a compendium of properties of spherical double-power-law models, some of which can also provide equally good matches to the density profiles of dark haloes. However, these models do not readily generalise to arbitrary distorted geometries. We show in our companion Paper II that the \rd{sNFW} model has another remarkable property. It can be used to form a new bi-orthonormal basis function expansion in a manner
analogous to that discovered by \citet{He92} for the Hernquist model
itself. Amongst other advantages, this means that the model can be
easily extended to flattened, triaxial and lopside geometries.}

 \rd{The intrinsic properties of our model (such as the DFs and velocity dispersions) are more complicated than the Hernquist model, but less complicated than the NFW model. We conclude that the sNFW model  provides an excellent trade-off between simplicity and realism in modelling dark haloes and elliptical galaxies.}

\section*{Acknowledgements}
EJL and JLS acknowledge financial support from the Science and
Technology Facilities Council. We thank Denis Erkal for providing us
with a suite of numerically constructed cosmological haloes, \rd{as well as the referees for constructive comments.}






\appendix

\section{Auxiliary Functions}

Here, we record some auxiliary functions used in the DF of the \rd{super-NFW} model. They are
\begin{eqnarray}
P_1(E) &=&-4 (32E^6\!+\!416E^5\!+\!1200E^4\!-\!920E^3\!-\!2198E^2\!+\!399E\!+\!504)\nonumber\\
P_2(E) &=&-8(32E^6\!+\!352E^5\!+\!656E^4\!-\!1176E^3\!-\!586E^2\!+\!173E\!+\!360)\nonumber\\
P_3(E) &=&(E\!+\!8)(128E^5\!+\!512E^4\!-\!576E^3\!-\!480E^2\!+\!56E\!+\!171)\nonumber
\end{eqnarray}


\bsp	
\label{lastpage}
\end{document}